\documentclass{article}
\usepackage[utf8]{inputenc}
\usepackage{spconf,amsmath,graphicx}
\usepackage{xcolor}
\usepackage{tablefootnote}
\usepackage{microtype}
\usepackage{cite}
\usepackage[hidelinks]{hyperref}
\usepackage{caption}
\usepackage{subcaption}

\title{Multichannel-based learning for audio object extraction}
\name{Daniel Arteaga \qquad Jordi Pons\thanks{\noindent The authors want to thank Giulio Cengarle for his insightful suggestions.}}
\address{Dolby Laboratories}

\begin{document}
\ninept
\maketitle
\begin{abstract}
    The current paradigm for creating and deploying immersive audio content is based on audio objects, which are composed of an audio track and position metadata.
    While rendering an object-based production into a multichannel mix is straightforward, the reverse process involves sound source separation and estimating the spatial trajectories of the extracted sources.
    Besides, cinematic object-based productions are often composed by dozens of simultaneous audio objects, which poses a scalability challenge for audio object extraction.
    Here, we propose a novel deep learning approach to object extraction that learns from the multichannel renders of object-based productions, instead of directly learning from the audio objects themselves. This approach allows tackling the object scalability challenge and also offers the possibility to formulate the problem in a supervised or an unsupervised fashion.
    Since, to our knowledge, no other works have previously addressed this topic, we first define the task and propose an evaluation methodology, and then discuss under what circumstances our methods outperform the proposed baselines.
\end{abstract}

\section{Introduction}
\label{sec:intro}

Audio objects, composed by an audio track and position metadata,
are rendered to specific listening layouts (e.g., 5.1 or stereo) during playback,
offering more flexibility, adaptability, and immersiveness than traditional multichannel productions.
The development of audio object extraction technologies allows for commercially interesting applications, such as converting old multichannel movies into audio formats based on objects---allowing them to get the most out of modern immersive reproduction venues, and paving the way for applications such as object modification or remapping.
However, object-based productions are composed by dozens of simultaneous audio objects. To overcome this object scalability challenge, several assumptions can be made. For example, some assume knowing (in advance) which sources are present in the multichannel mix. Accordingly, source-specific models can be trained to extract such objects~\cite{lluis2018end,stoter19,kavalerov2019universal}.

This approach presents the difficulty that in cinematic object-based productions all kinds of audio objects can be present in the mix, making it difficult to assume (in advance) which sources will be present.
Another direction relies on the assumption that a fixed set (e.g., 1 or 10) of unknown objects are present in the mix. While these models are not source-specific,
 the number of audio objects in a realistic production is often large and those are limited in the amount of simultaneous objects they can extract~\cite{roebel2015automatic,salaun2014flexible,mitsufuji2014online}. 

\begin{figure}[t]
    \centering
    \includegraphics[width=\columnwidth]{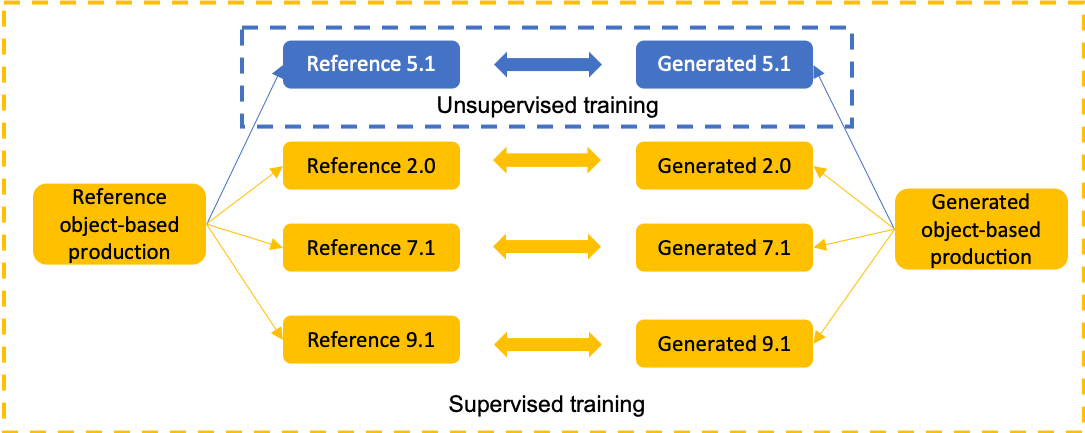}
    \caption{Multichannel-based learning: supervised and unsupervised. For supervised learning, a set of multichannel mixes, rendered from a reference object-based production, are compared. In contrast, unsupervised learning only relies on the 5.1 mixes and does not require a reference object-based production.} 
    \label{fig:training_configurations}
    \vspace{-6mm}
\end{figure}

Such object scalability issue also poses challenges from the model optimization perspective. It is  challenging to define the reference objects for object-based supervised learning. 
The problem of permutation ambiguity, described in the speech~\cite{kadioglu2020empirical,luo2018tasnet,liu2019divide} and universal~\cite{kavalerov2019universal,wisdom2020unsupervised} source separation literature, also arises here. The output to ground truth pairs required for supervised learning can not be arbitrarily assigned due to the source- (or speaker-) independent nature of the task.
Note that in cinematic audio productions, the number of potential object permutations can be very large, since these typically contain dozens of simultaneous objects. For this reason, permutation invariant training, used for speech and universal source separation,  is impractical.
Aggravating the problem, since audio engineers can combine several objects to present an audio event, some of the reference objects may only correspond to parts of real auditory events.

To overcome these challenges, we propose an approach based on multichannel-based learning: our references for supervised learning are not objects, but rather the multichannel mixes rendered from those objects. Multichannel-based learning is inspired by the way humans evaluate cinematic productions, by which two object-based productions are considered to be similar if their renderings to multichannel layouts are also similar---even though the number of objects in the two mixes might be different. We propose to extract (i) a small number of objects, typically 1--3, corresponding to the most prominent auditory events; and (ii)~a multichannel remainder, called  ``bed channels", containing the audio not embedded in the objects. Hence, we study source-independent deep learning models extracting up to 3 objects, in addition to the bed channels.

Unsupervised learning has also been useful to extract a fixed set of unknown objects from a mix. For example, low-rank approximation techniques, such as non-negative matrix factorization, have been quite successful at estimating a few object sources from a given mix~\cite{roebel2015automatic,salaun2014flexible,mitsufuji2014online}. 
Given the success of unsupervised methods in the past, we also explore an unsupervised version of the multichannel-based deep learning approach we propose. 
Our model relies on inductive biases in the form of loss and architectural constraints to facilitate disentangling audio objects, including position metadata, from the mix \cite{locatello2019challenging}. Our loss constraints impose a set of desired properties for the objects and our architecture constraints rely on a fully-differentiable (but not learnable) renderer that converts the estimated objects and position metadata into a multichannel mix. As we further discuss in Sect.~2, the renderer (a fully-differentiable implementation~\cite{engel2019ddsp} of Dolby Atmos \cite{thomas2017amplitude}) enables multichannel-based learning, as opposed to object-based learning, because it dictates a fully-determined object-based format at the encoder's output.   
Finally, note that multichannel-based learning also bears resemblances with self-supervised learning \cite{ravanelli2020multi,pascual2019learning}: instead of directly learning from objects, our loss targets a proxy signal based on multichannel renders.

In the following, we introduce supervised and unsupervised multichannel-based learning (section 2), explain the inductive biases we employ on both the architecture \mbox{(section 3)} and on the loss function (section 4), and introduce a novel evaluation methodology based on Freesound Datasets data (section 5).

\section{multichannel learning}
\label{sec:training}

\begin{figure}[t]
	\centering
	\includegraphics[width=1\columnwidth]{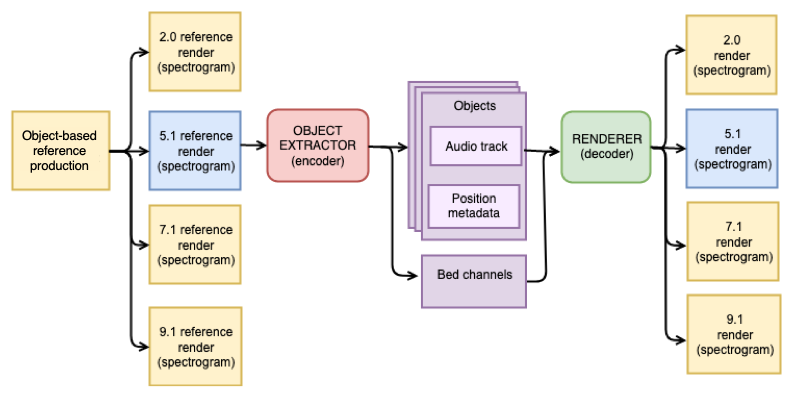}
	\caption{The multichannel-based audio object extraction model. For both inference and training the learnable object extractor (encoder, see Fig.~\ref{fig:diagram_zoom}) extracts the objects and bed channels out of the input 5.1. For training, the non-trainable differentiable renderer decodes them to a number of layouts. For unsupervised training, the objective function is based on the 5.1 mixes only (blue boxes). For supervised training, other renders are additionally taken into account (blue and yellow boxes).}
	\label{fig:diagram}
\end{figure}

\begin{figure}[t]
	\centering
	\includegraphics[width=1\columnwidth]{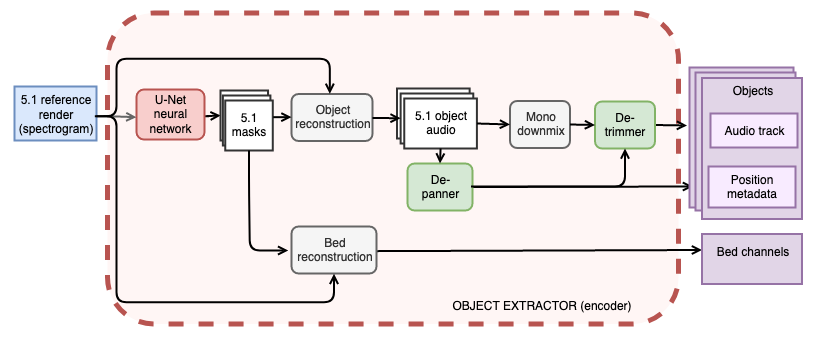}
	\caption{The encoder, which extracts the  objects and bed channels from a 5.1 mix. As in in Fig.~\ref{fig:diagram}, the red box denotes the learnable part of the model, and green boxes represent the non-learnable differentiable digital signal processing parts.}
	\label{fig:diagram_zoom}
\end{figure}

We design and build a neural network which, given a multichannel excerpt,  extracts a fixed number of audio objects, position metadata, and a multichannel reminder (the bed channels). For simplicity, we  assume a 5.1 input---although our method can be extended to any other multichannel input format. As introduced above, our training objective does not rely on object-based supervised learning. Instead, our training objective is designed to learn from multichannel renders in a supervised or unsupervised fashion (see Fig.~\ref{fig:diagram}):
\vspace{4mm}
\begin{itemize}
    \item \textbf{Supervised learning}. An object-based reference mix is required to render a set of pre-defined multichannel layouts (e.g.,~2.0, 5.1, 7.1, 9.1). 
    The obtained renders are used as reference for a reconstruction loss that is defined in the multichannel layout domain. This loss consists on a weighted average of the reconstruction losses obtained for each multichannel format.
    This supervised configuration requires a training set of object-based productions, from which all reference multichannel renderings are derived. 
    
    \item \textbf{Unsupervised learning}: the unsupervised configuration can be understood from the perspective of an autoencoder, where the 5.1 input is encoded into a pre-defined latent space (the space of object-based productions, see Fig.~\ref{fig:diagram}), and decoded to reconstruct the 5.1 input. Hence, a training set of object-based productions is not required, since the unsupervised reconstruction loss is just defined over 5.1 signals.
\end{itemize}

Multichannel-based learning is enabled by the structure and the inductive biases of the model. For example, the neural network we employ dictates a fully-determined object-based format in the latent space---including, among others, the number of objects, the presence of a bed channel, and the specifics of the renderer (see Figs.~\ref{fig:diagram} and~\ref{fig:diagram_zoom}).
Furthermore, the position, motion and object content of the extracted objects are further constrained by penalties added to the loss function (see Sect.~\ref{sec:loss}). These penalties encourage the extracted objects to have an independent semantic meaning and to behave consistently with how an audio engineer would mix the objects.  
Hence, multichannel-based learning is enabled by the structured way we approach the problem. We enforce such structure via the architecture of the model, and via additional regularization terms in the loss function.

Note that unsupervised learning could also be used to fit a specific multichannel excerpt. Fitting a specific 5.1 mix via unsupervised learning enables audio object extraction without requiring any training database. This ``unsupervised fit" case can be seen as a 5.1 to 5.1 autoencoder that overfits a specific example, where the structure of the model and the guidance of the regularization loss terms tailor the model towards extracting meaningful audio objects in the latent. Due to the intriguing properties of the ``unsupervised fit" case, our unsupervised learning experiments focus on understanding the viability of the ``unsupervised fit" approach.  Furthermore, supervised and ``unsupervised fit" learning can be combined: a model can be pre-trained with an object-based dataset via supervised learning, which can be fine-tuned to a specific 5.1 excerpt with ``unsupervised fit" training. We refer to this combined configuration as ``fine-tuned".

	\vspace{-2mm}

\section{MODEL ARCHITECTURE}
\label{sec:architecture}

	\vspace{-1mm}

Our model consists of an object extractor module (encoder) and a renderer module (decoder)---see Fig.~\ref{fig:diagram}. The encoder (Fig.~\ref{fig:diagram_zoom}) performs audio object extraction and converts the 5.1 input into an object-based format. The goal of the decoder is to render the extracted objects and bed channels into multichannel mixes to allow supervised or unsupervised multichannel-based learning.

The encoder (Fig.~\ref{fig:diagram_zoom}) is composed of
(i) the mask-estimation block, a trainable deep neural network that estimates object and bed channel masks; (ii) the rest of object-extraction blocks, which extract the audio objects (including position metadata) and bed channels out of the estimated masks. The mask-estimation block (i) operates over a 5.1 mel power spectrogram excerpt and extracts $n$ object masks and a bed channel mask. Our source separation model is based on U-Net \cite{Jansson2017SingingVS}.
However, note that our method could also be extended to waveform-based models such as Wavenet~\cite{lluis2018end,rethage2018wavenet} or Wave-U-Net~\cite{stoller2018wave,defossez2019music}. The object-extraction blocks (ii) rely on differentiable digital signal processing layers to further process the object masks and bed channels masks for reconstructing the objects and bed channels.  Important blocks among them are the de-panner and the de-trimmer. The de-panner extracts the position metadata from the 5.1 object audio. Our current implementation is based on a differentiable digital signal processing layer; however, it could also be extended to be a learnable deep neural network. The de-trimer reverses the ``trimming'' process (the reduction of the object level for objects far from the frontal position, applied during rendering).
The decoder is just a fully-differentiable audio-object renderer, which renders the object and bed channels to specific multichannel layouts. While all the layers of our model are differentiable, only the mask-estimation block (i)  is trainable.

For inference, only the encoder is used: the output of the model are the $n$ objects (including audio and spatial metadata) and the bed channels estimated by the encoder.  For training, the objects and bed channels out of the encoder are rendered by the decoder to a set of different layouts (2.0,~5.1,~7.1~and~9.1 for supervised learning, and 5.1 for unsupervised learning) to compute the reconstruction loss.

In our implementation, the entire model is written in Tensorflow, including both the trainable and non-trainable digital signal processing modules. The mask-estimation block implements the U-Net \cite{Jansson2017SingingVS}, with minimal adaptations in the final layer to generate 5.1 masks. The renderer is the Dolby Atmos renderer \cite{thomas2017amplitude} expressed in differentiable form. The de-panner and de-trimmer also correspond to the Dolby Atmos renderer. The model operates on audio excerpts of 5.44 seconds at 48$\,$kHz, with an FFT window length of 2048 samples, leading to audio patches of 256 time bins and 1025 frequency bands, which are grouped in 128 mel bands.

\section{Training objectives}
\label{sec:loss}

We rely on two main training objectives: reconstruction losses, to match the content of the mix at the multichannel level, and regularization losses, a set of penalties that encourage the extracted objects to behave consistently.

\textbf{Reconstruction losses} --- These are derived from the comparison between the reference renders/mixes and the outputs of the decoder. As discussed in Sects.~\ref{sec:training} and \ref{sec:architecture}, in multichannel-based supervised learning we compare several reference renders (2.0, 5.1, 7.1 and 9.1, rendered from the reference object based production) to the corresponding decoder outputs. Whereas for multichannel-based unsupervised learning, we compare the reference 5.1 mix to the 5.1 output of the decoder (see Fig.~\ref{fig:training_configurations}). We got our best results when using the $L_1$ loss. 
For supervised learning, the resulting reconstruction loss is computed by a weighted average of the losses obtained for each multichannel layout. We give a higher weight to more dense layouts like 7.1 and 9.1, because they provide more resolution and give better results in practice. 

\textbf{Regularization losses} --- To motivate the need of these regularization terms, notice that, for the unsupervised case, the model could trivially minimize the reconstruction loss to zero by sending all content to the bed channels.  
In either the supervised or unsupervised cases, the network can learn to minimize the cost function in ways that result in  extracted objects that are not usable---e.g., having objects moving from one position to another distant position without any continuity, or having static objects close to the speakers in a 5.1 layout (behaving like bed channels).
It is therefore necessary to bias the model towards solutions which correspond to the way object-based productions are expected to be. We found that a convenient way to do so is via regularization loss terms. 
The regularization loss terms we consider are  the following: (i) bed channel content, to avoid trivial solutions and encourage object creation; (ii) objects close to the 5.1 loudspeaker positions, to discourage objects playing the role of bed channels; (iii) very slowly moving objects, to limit the amount of static objects; (iv) very rapidly accelerating objects, to avoid ``jumpy'' objects moving instantaneously from one side of the room to the other; (v) objects close one to each other;  (vi) correlated object trajectories; and (vii) correlated content among objects, or between objects and bed channels, to avoid  objects and bed channels sharing the same content. The relative weight of each one of the different loss penalties corresponds to tunable hyperparemeters of the model. By tuning the different loss weights, the model can be made to behave more or less aggressively at extracting objects. The scale of the different regularization loss terms is between 0.1\% and 100\% of the reconstruction loss magnitude approximately. 
We found useful to apply the largest penalization to terms (i) and (iv).

\begin{figure}
	
	\centering
	\includegraphics[width=0.47\columnwidth]{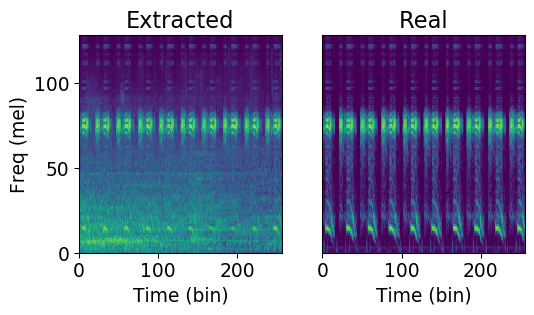}
	\centering
	\includegraphics[width=0.55\columnwidth]{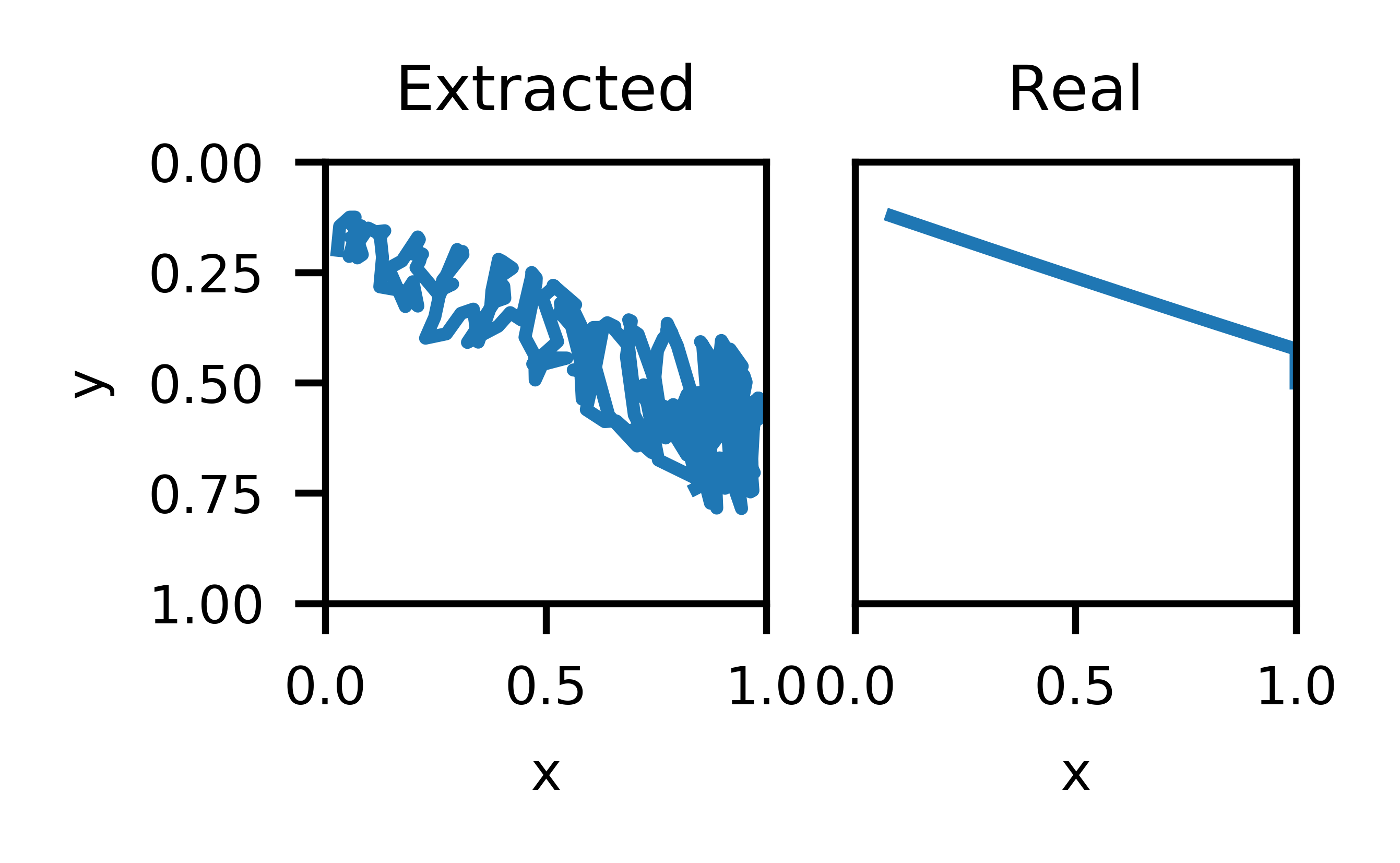}
	\caption{Example object extracted by the multichannel-based learning model and corresponding reference. Model configured to extract one object and trained in supervised mode. Top: object audio (spectrograms); bottom:  object position metadata ($xy$ trajectories).The input 5.1 and output bed channels are shown in Fig.~\ref{fig:example-spectrogram-expanded} in the Appendix.}\label{fig:example-spectrogram}
		\vspace{-3mm}
\end{figure}

\section{EXPERIMENT AND EVALUATION}
\label{sec:evaluation}

\subsection{Experiment methodology}

\noindent Since we are not aware of a previously established methodology to evaluate object extraction methods, we propose two experiments, differing in the number of  objects to be extracted (1 or 3). The task amounts to extracting 1 or 3 objects and the bed channels from a 5.1 mix, rendered from known objects that are available for evaluation. 
The first experiment consists in extracting objects from fifty 5.1 mixes containing 1 object and bed channels, and the second experiment consists in extracting objects from another set of fifty 5.1 mixes having 3-objects + bed channels. The objects in the mixes have been created by assigning pseudo-random synthetic trajectories to real audio tracks representative of different sound categories appearing in cinematic mixes (vehicle sound, special effects, music instruments, voices, footsteps, etc), obtained from Freesound Datasets~\cite{fonseca2017freesound,fonseca2018general,fonseca2020fsd50k}. These object-based excerpts also contain bed channels with real surround recordings. All objects and bed channels in a given excerpt are set to the same  level. We  use  the 1-object and 3-object excerpts to evaluate two object extraction systems capable of extracting 1 and 3 objects, respectively.
Example of the real spectrograms and trajectories, and the result of the inferred signals by the 1-object model, are shown in Fig.~\ref{fig:example-spectrogram}.
To evaluate the performance of the object and bed channels extraction we compute the scale-invariant signal-to-distortion ratio (SI-SDR) \cite{leroux2019sdr} of the extracted objects with respect to the ground truth audio objects, with the goal of evaluating how well multichannel-based learning methods can extract individual objects. Since the order of the extracted objects is arbitrary, in the 3-object case we compute all possible permutations between the extracted objects and consider the permutation which gives the best median SI-SDR value considering the three objects.

As a baseline for the object extraction evaluation, we consider a naive mono downmix $(\mathrm{L} + \mathrm{R} + \mathrm{C} + \mathrm{Ls} + \mathrm{Rs})/5$ for the 1-object experiment, and a downmix to a $(2\mathrm{L} + \mathrm{C})/3$, $(2\mathrm{R} + \mathrm{C})/3$ and $(\mathrm{Ls} + \mathrm{Rs})/2$ for the 3-object experiment. These baselines correspond to one of the best possible methods to extract distinct objects in absence of source separation techniques. For the bed channels, the baseline consists in a bypass of the input 5.1 directly to the bed channels. In Figs.~\ref{fig:results-1}--\ref{fig:bed-3} we depict the SI-SDR improvement (SI-SDRi) with respect to the baseline (the optimal permutation of the 3-channel downmix baseline is also considered). As an upper reference, we consider the ideal binary mask (IBM). To assess the impact of the mel scale masks, we compute the IBM with and without frequency mel band grouping.

We evaluate the performance of each model in three different training configurations: ``supervised", ``unsupervised fit", and ``fine-tuned" (see Sect.~\ref{sec:training}). 
For the the supervised and fine-tuned training configurations, the models are trained, or pre-trained, with object-based excerpts generated on the fly, created by pseudo-random  procedurally-generated audio samples and synthetic trajectories, amounting to a virtually infinite training and evaluation datasets.   The model was trained using the Adam optimizer \cite{adam} with a learning rate of $2\times10^{-4}$ and a batch size of 8.

\subsection{Results}

\begin{figure}
    \centering
         \vspace{-10mm}

    \includegraphics[width=0.6\columnwidth]{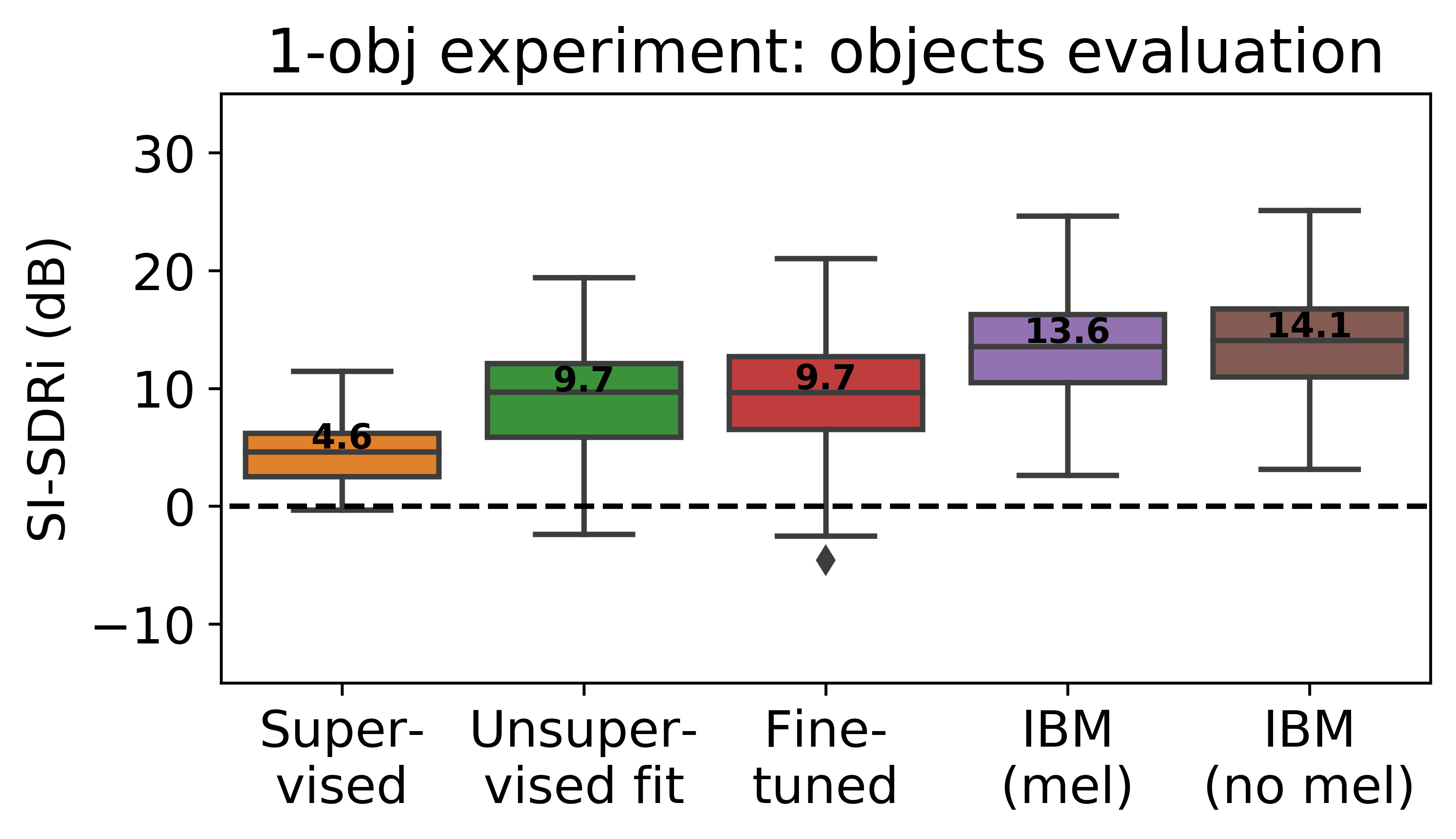}
                \vspace{-3mm}

    \caption{Object extraction evaluation: 1-object experiment.}\label{fig:results-1}
\end{figure}

\begin{figure}
    \centering
             \vspace{-2mm}

    \includegraphics[width=0.6\columnwidth]{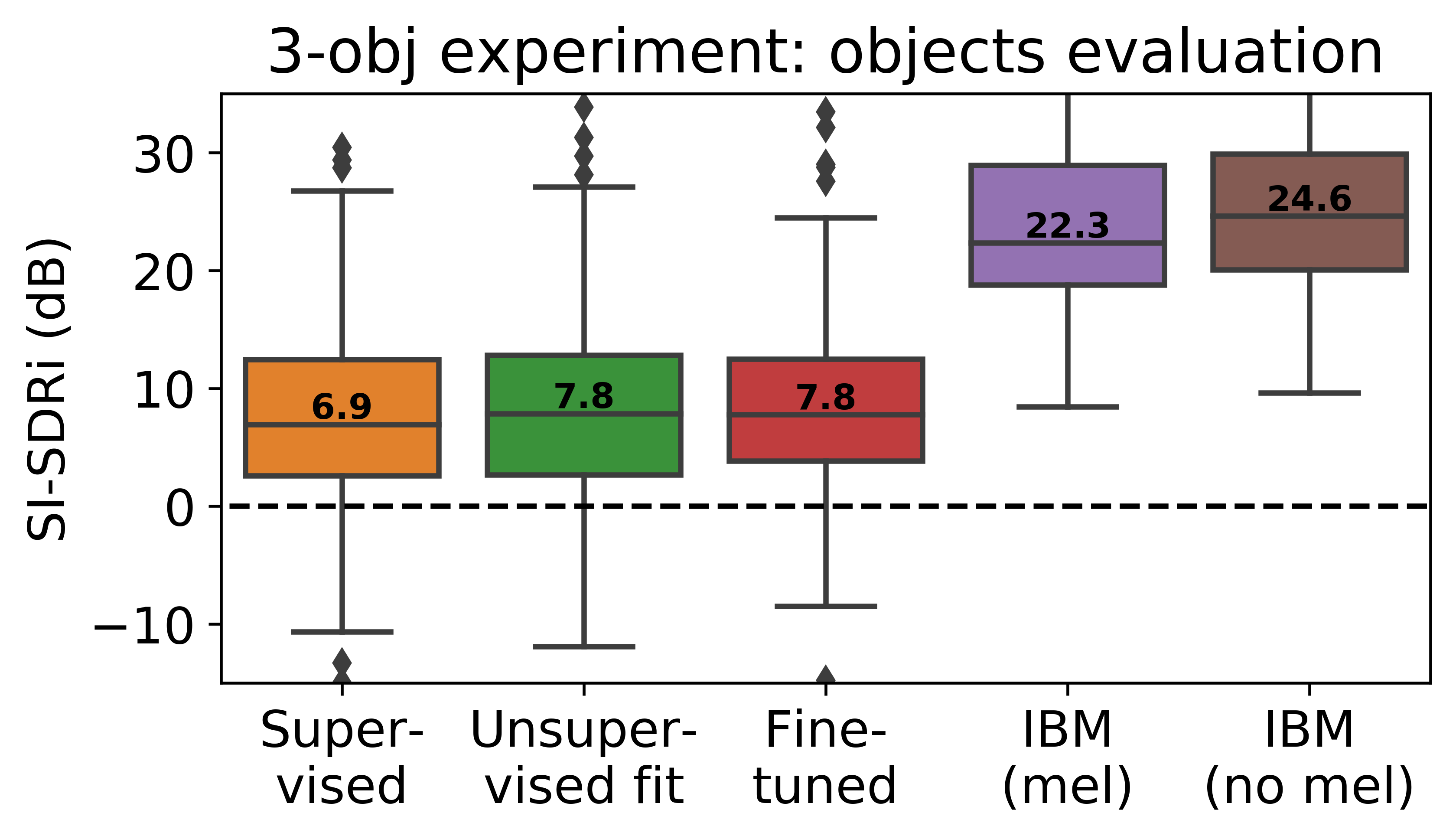}
                \vspace{-3mm}

    \caption{Object extraction evaluation: 3-object experiment.}\label{fig:results-3}
    \vspace{-4.2mm}
\end{figure}

\begin{figure}[t]
    \centering

     \vspace{-10mm}

    \includegraphics[width=0.6\columnwidth]{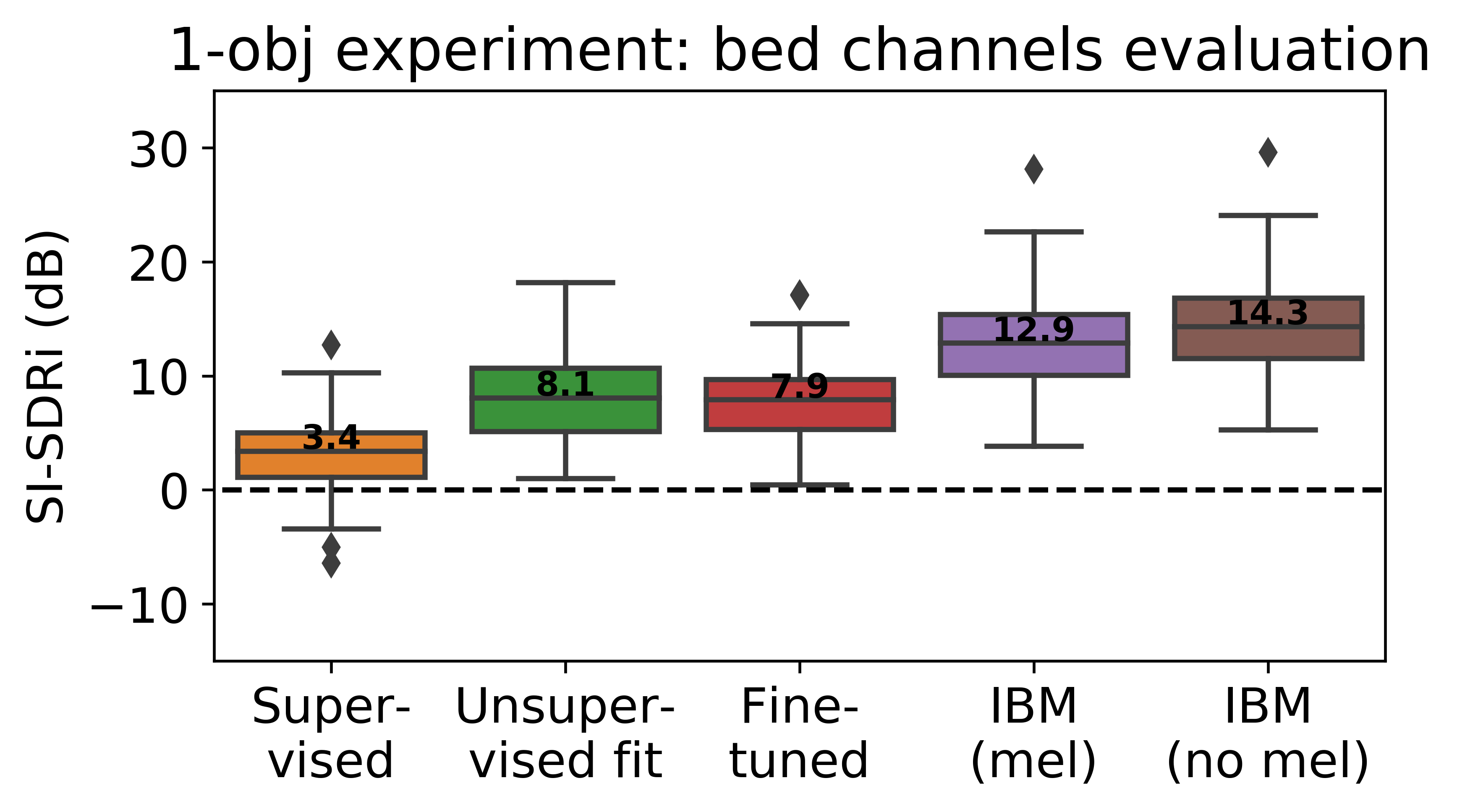}
                \vspace{-3mm}

    \caption{Bed channels extraction evaluation: 1-object experiment. }\label{fig:bed-1}
\end{figure}

\begin{figure}[t]
    \centering
        \vspace{-2mm}
    \vspace{0.9mm}
    \includegraphics[width=0.6\columnwidth]{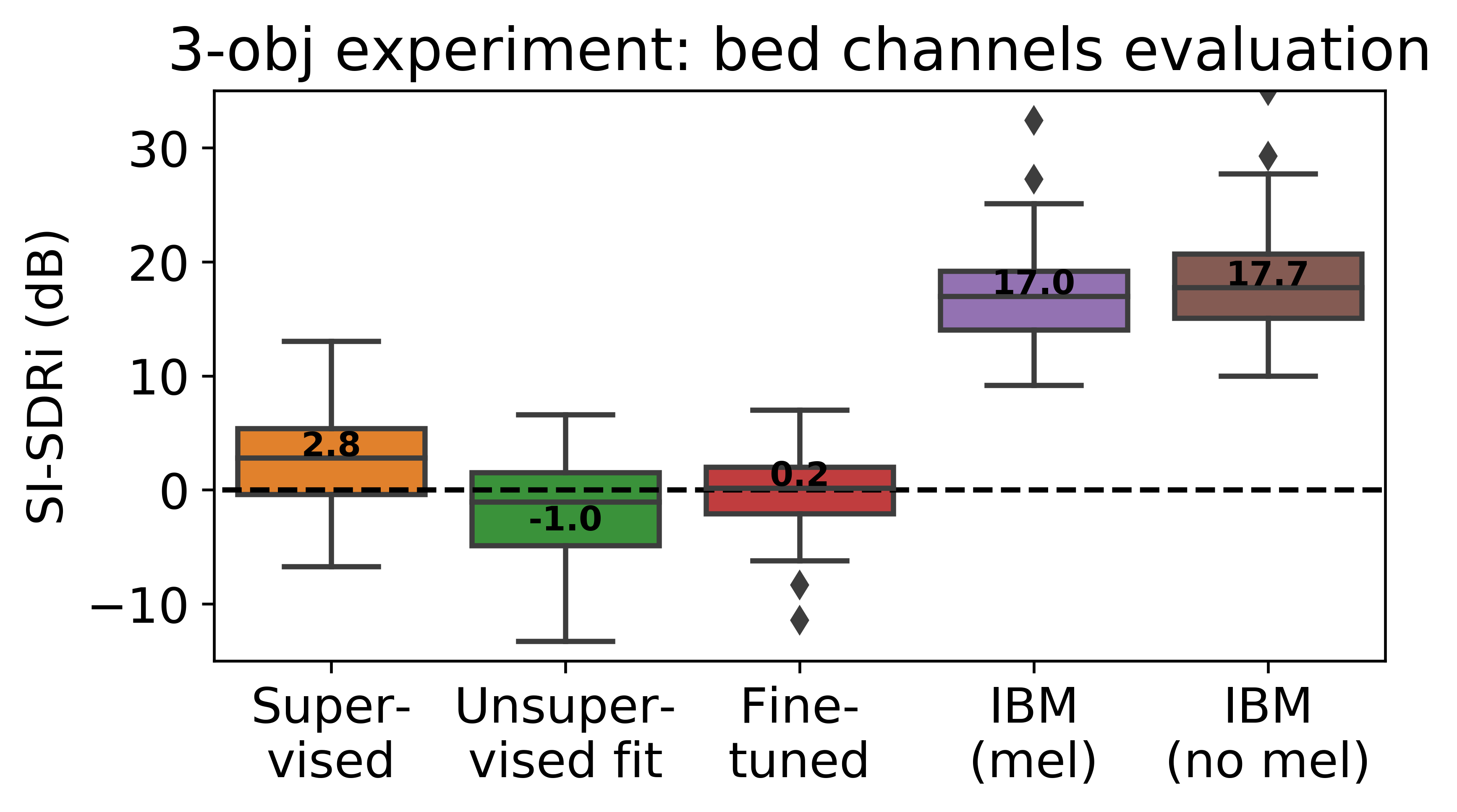}
            \vspace{-3mm}
    \caption{Bed channels extraction evaluation: 3-object experiment.}\label{fig:bed-3}
    \vspace{-3.8mm}
\end{figure}

SI-SDRi object extraction results are reported in Figs.~\ref{fig:results-1}, \ref{fig:results-3}. 
For all three training configurations, both 1- and 3-object models showcase a noticeable improvement with respect to the baseline (dashed line at 0 dB). Regarding the 1-object experiment, supervised training {gives} modest SI-SDR improvements, but also more {robust} (with a smaller data variability).  Further, ``unsupervised fit" and ``fine-tuned" (which explicitly overfit the 5.1 excerpt) deliver a notable performance, rather close to the ideal mask results, indicating that most of the times the model is able to estimate the right object trajectory and to extract its corresponding audio object. 
Regarding the 3-object experiment, there is also a noticeable improvement with respect to the baseline. However, our results are far from the IBM results, denoting the difficulty of separating three objects from a complex and dense mix.
Informal listening reveals that, sometimes, under challenging conditions, the extracted objects tend to follow a given reference object for just some time, after which such extracted object changes to track another audio source. Additionally the extracted objects do not include strong separation artifacts. This can be attrubuted to the fact tha our model is based on spectral masks (filtering): it cannot generate any sound which is not already present in the original multichannel mix; the only possible artifacts are those associated to the time-frequency masking.

While in most cases the main interest is in evaluating the extracted objects, it is also interesting to evaluate the quality of the bed channels (see Figs.~\ref{fig:bed-1}, \ref{fig:bed-3}). In our 1-object experiments the extracted bed channels clearly outperform the baseline. However, in our 3-object experiments, the ``unsupervised fit" and ``fine-tuned" configurations cannot outperform the baseline. This result is illustrative, because it reveals the strengths and weaknesses of the ``unsupervised fit" and ``fine-tuned" approaches. While these achieve among the best results at object extration via overfitting a specific 5.1 excerpt, the strong inductive biases we introduce (through the regularization losses and the architecture) result in an aggressive object extraction that can compromise the quality of the bed channel. This is particularly noticeable for the ``unsupervised fit" case, that is trained from scratch to overfit a given 5.1 without additional training data.

\section{CONCLUSIONS AND DISCUSSION}
\label{sec:conclusions}

We proposed a source-independent approach relying on strong inductive biases to learn from multichannel renders. The inductive biases we explore are based on architectural constraints (enforcing the bottleneck of our model to be a specific object-based format), and on additional regularization loss terms (enforcing objects to behave according to object-based production conventions).
Multichannel-based learning can be formulated in a supervised or unsupervised fashion. We found that supervised multichannel-based learning delivers solid results, extracting useful and congruent objects, even if the model was optimized to approximate multichannel mixtures, not to extract audio objects.. Interestingly, we also found that unsupervised multichannel-based learning can improve the results obtained by supervised multichannel-based learning.
Specifically, we looked at the ``unsupervised fit" configuration, which trains the model from scratch to overfit a given 5.1 without additional training data. While ``unsupervised fit" results are among the best, its separations can be aggressive, harming the quality of the bed channels. Furthermore, it is slow since the model needs to be optimized for any given example.
Finally, we also found that combining supervised pre-training with ``unsupervised fit" delivers results that are comparable to ``unsupervised fit" alone.
	
\bibliographystyle{IEEEbib}
\bibliography{refs}

\newpage
\appendix
\section{Example results}

\begin{figure}[h]
    \centering
    \begin{subfigure}[b]{\columnwidth}
        \centering
        \includegraphics[width=0.7\columnwidth]{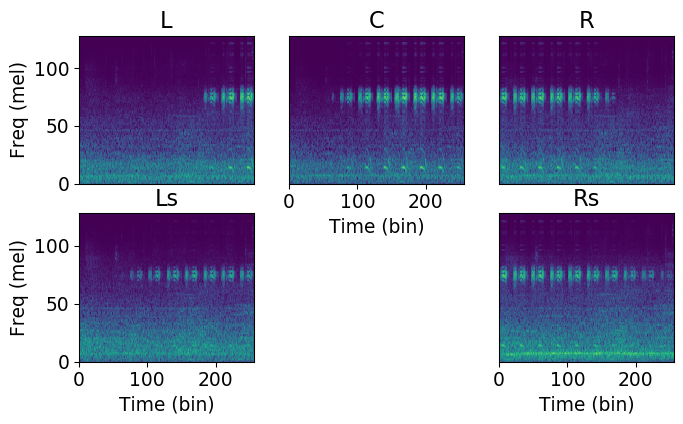}
        \caption{Input: 5.1 (spectrogram).}
    \end{subfigure}
    \begin{subfigure}[b]{\columnwidth}
        \centering
        \includegraphics[width=0.6\columnwidth]{supervised_obj_0.png}
        \caption{Output: object audio (spectrogram, compared to reference).}
    \end{subfigure}
    \begin{subfigure}[b]{\columnwidth}
        \centering
        \includegraphics[width=0.65\columnwidth]{traject.png}
        \caption{Output: object position metadata ($xy$ trajectory, compared to reference).}
    \end{subfigure}
    \begin{subfigure}[b]{\columnwidth}
        \centering
        \includegraphics[width=0.7\columnwidth]{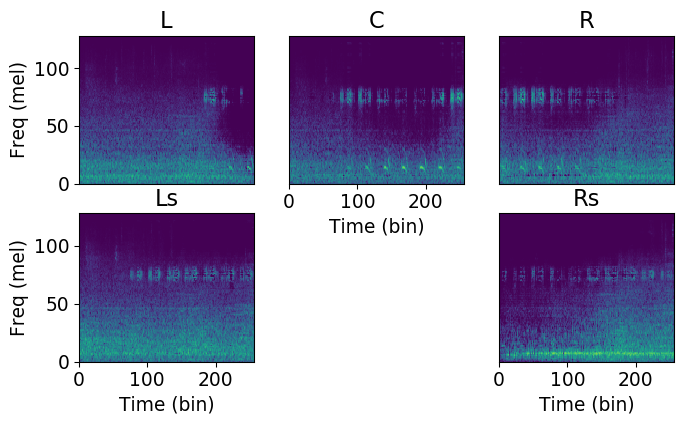}
        \caption{Output: bed channel residual (spectrogram).}
    \end{subfigure}
   \caption{Example inputs and outputs of the multichannel-based learning model, configured to extract one object and trained in supervised mode. For the 5.1 spectrograms only the 5.0 channels are depicted.}
    \label{fig:example-spectrogram-expanded}
\end{figure}

Figure~\ref{fig:example-spectrogram-expanded} below is an expanded version of  Fig.~\ref{fig:example-spectrogram}.

\end{document}